\documentclass[aps,prl,twocolumn,groupedaddress]{revtex4}
\usepackage{graphicx}

\begin{document}


\title{Six-fold configurational anisotropy and magnetic reversal in nanoscale Permalloy triangles }

\author{L. Thevenard} \email{l.thevenard@imperial.ac.uk}
\author{H. T. Zeng}
\author{D. Petit}
\author{R. P. Cowburn}
\author{}

\affiliation{Blackett Laboratory, Physics Department, Imperial College London, Prince Consort Road, SW7 2BW London, United Kingdom}


\date{\today}

\begin{abstract}

Six-fold configurational anisotropy was studied in Permalloy triangles, in which the shape symmetry order
 yields two energetically non-degenerate micromagnetic configurations of the spins, the so-called "Y" and "buckle" states. A twelve pointed switching astroid was measured using magneto-optical experiments and successfully reproduced numerically, with different polar quadrants identified as specific magnetic transitions, thereby giving a comprehensive view of the magnetic reversal in these structures. A detailed analysis highlighted the necessity to include the physical rounding of the structures  in the simulations to account for the instability  of the Y state.

\end{abstract}


\maketitle

\begin{figure}
	\centering
					\includegraphics[width=0.5\textwidth]{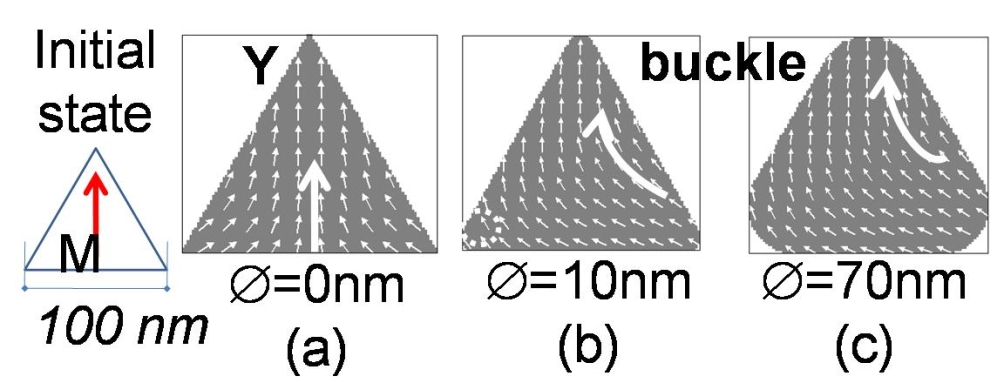}
		
		\caption{Micromagnetic simulations at equilibrium after a Y state initialization (thickness 10~nm, triangle base $e=100$~nm). (a) In sharp triangles, the Y state is stable for $e\leq 200$~nm. (b) and (c) As soon as a finite rounding is included, the spins preferentially slide toward to a buckle state.}
	\label{fig:rounding}
\end{figure}

The growing necessity for ultra-high density data storage has spurred the exploration of different schemes based on magnetic media \cite{allwood02,parkin08,pappert07}. In most cases, the challenge when scaling structures down is to keep the magnetization stable against thermal fluctuations by controlling its magnetic anisotropy, while keeping accessible read/write fields or currents for the device \cite{cowburn03}. The detailed understanding of magnetic reversal mechanisms in magnetic nanostructures is therefore essential, as well as novel ways to control magnetic anisotropy. This can be done by material engineering (tuning the magneto-crystalline anisotropy), or by nanostructure engineering, using shape anisotropy in soft ferromagnets. In thin elliptical Permalloy (Py) nanoelements for instance, shape anisotropy forces the magnetization to lie along the longer axis. In higher-order symmetry elements, an additional phenomenon called configurational anisotropy (CA) appears, in which the high demagnetization energy cost of a uniform magnetization is such that the spins rearrange themselves to yield radically different micromagnetic configurations. First explored by Schabes and Bertram \cite{schabes88}, it can be seen as a higher-order form of shape anisotropy. In triangles for example, the magnetization either fans in from two corners toward the third with $\vec{M}$ along the bisector ("Y" state, Fig. \ref{fig:rounding}a), or bends toward one of the corners with $\vec{M}$  parallel to the edge ("buckle" state, Fig. \ref{fig:rounding}b-c) \cite{koltsov00}. CA has also been demonstrated experimentally in cubes \cite{schabes88}, squares and pentagons \cite{cowburn99,vavassori}. 

In this letter, we relate this behavior in Py triangles with a detailed understanding of the micromagnetic reversal mechanisms, by combined magneto-optical measurements and micromagnetic simulations (\textsc{OOMMF} package  \endnote{OOMMF code available at http://math.nist.gov/oommf.}). Great attention was given to the rounding of the triangles chosen for the simulation bitmaps, and a  twelve-pointed switching astroid was found both experimentally and numerically,  providing a hitherto lacking tool for a complete description of CA in ferromagnetic triangles.

\begin{figure}
	\centering
								\includegraphics[width=0.35\textwidth]{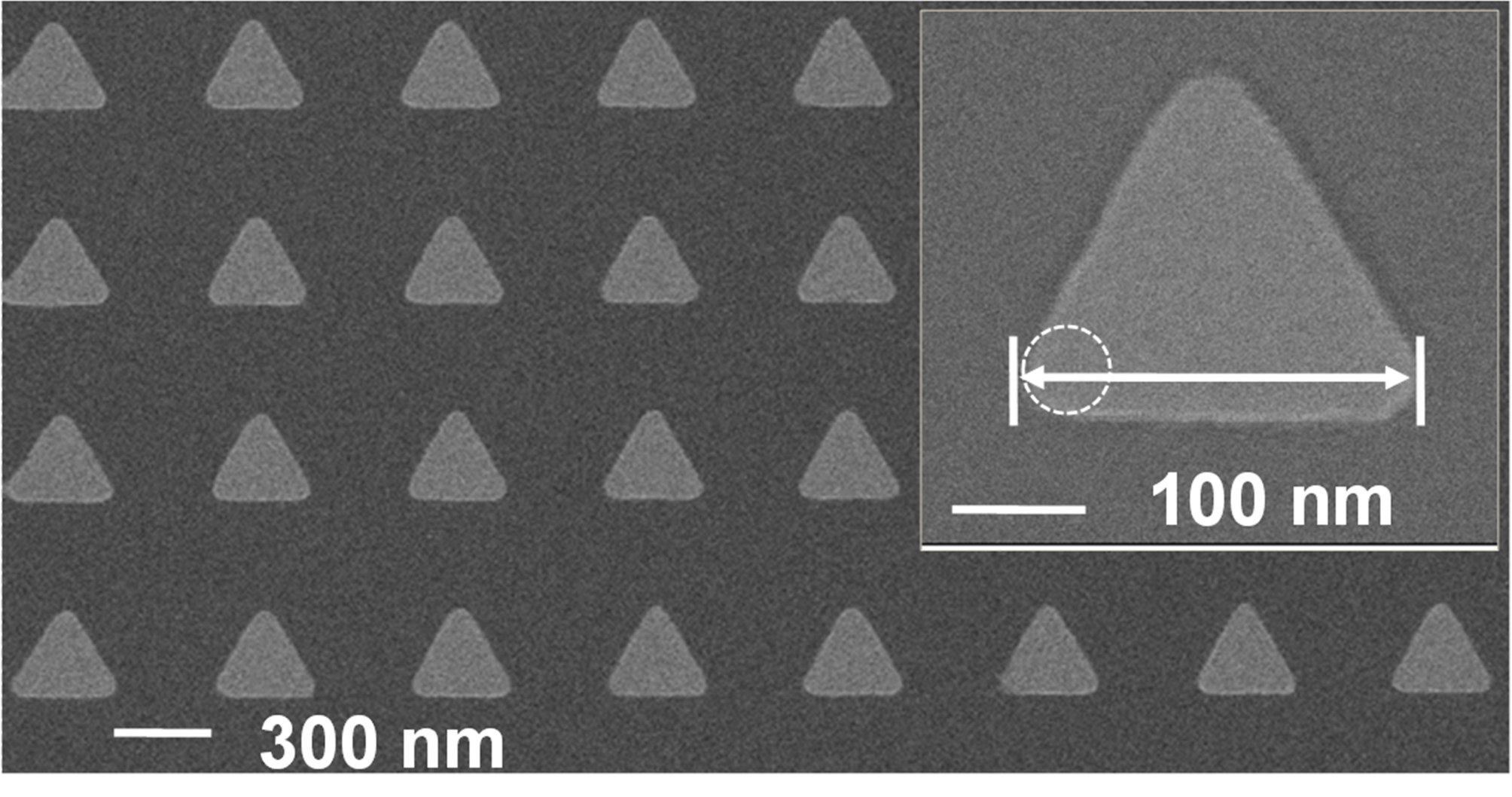}
				
			\caption{Scanning electron micrograph of 300~nm wide, 9~nm thick Py triangles, spaced by 300~nm and presenting an average rounding of 70~nm.}
	\label{fig:SEMarray}
\end{figure}

Large $25\times25\mu m^{2}$ arrays of 9~nm thick Py equilateral triangles were fabricated on Silicon by a 20kV electron beam lithography, followed by thermal evaporation and a lift-off process. The triangles had a $e$=300~nm base and were spaced by $e$, for which magnetostatic coupling between elements is negligible. The average corner rounding was $\oslash = 70$~nm (Fig. \ref{fig:SEMarray}), partly inherent to the fairly low electron energy used in the lithography step.

\begin{figure}
	\centering
\includegraphics[width=0.4\textwidth]{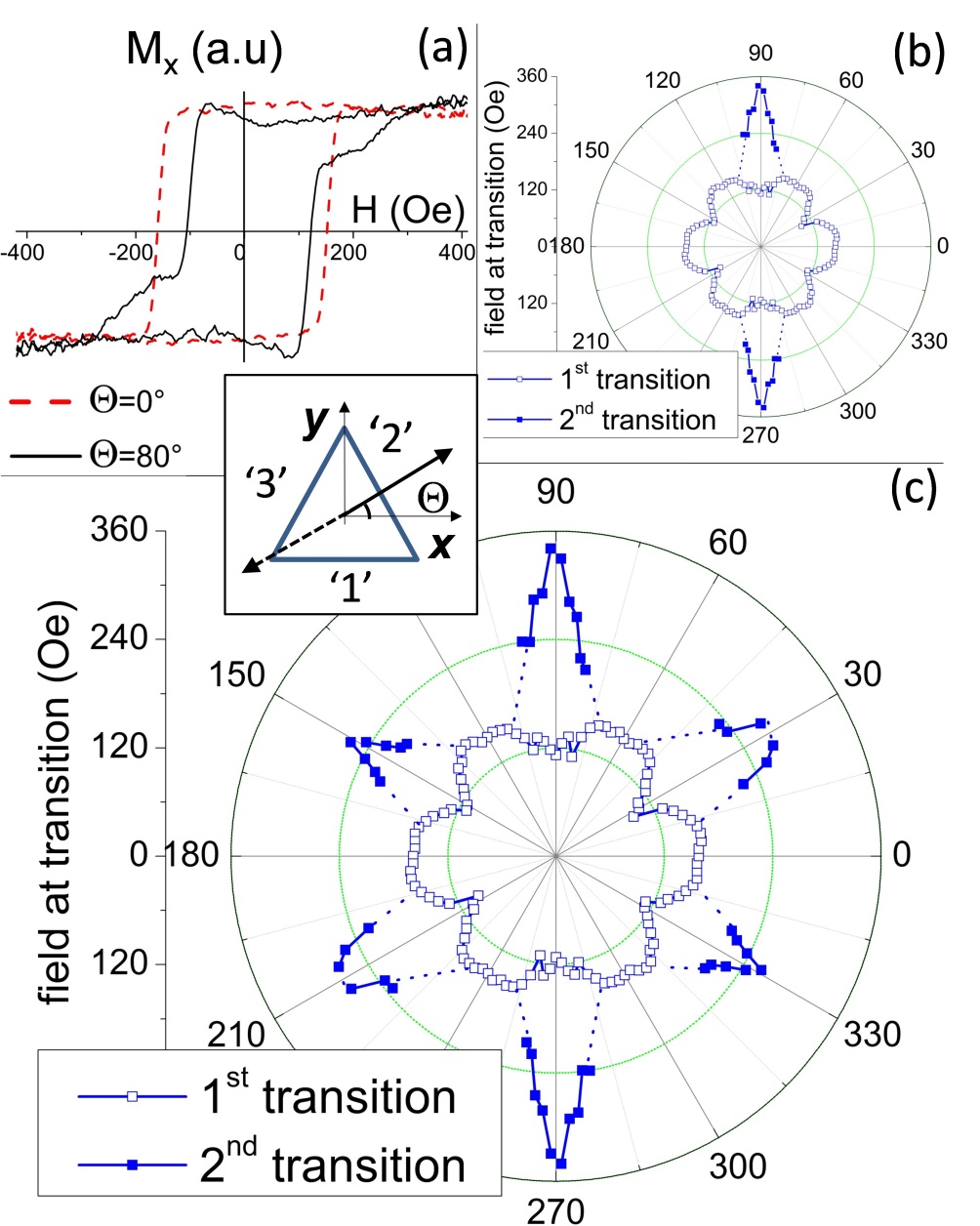}

	\caption{(color online) (a) Longitudinal Kerr magnetization along edge '1': single transition for $\Theta = 0^{\circ}$ (dashed line), double transitions for $\Theta = 80^{\circ}$(solid line). (b) Data points obtained with the Kerr axis along edge '1': amplitude of the field at transition plotted against its direction. Open (closed) symbols correspond to first (second) transitions. (c) Full switching astroid. The dotted lines are guides to the eye.}
	\label{fig:astroid}
\end{figure}

We then used a highly sensitive magneto-optic Kerr effect magnetometer (MOKE) to probe the longitudinal component of the magnetization. The magnetic field could be applied along any in-plane direction as a combination of ($H_{x}$,$H_{y}$) fields created by a quadrupole electromagnet, and the laser  was focused in a $5~\mu m$ spot on the array. The coercivity of the unpatterned film was $H_{c}$=2~Oe, and its  magneto-crystalline anisotropy field $H_{an}$=14~Oe, measured on macroscopic structures on the sample (along $x$ for $H_{c}$, and $y$ for $H_{an}$, Fig. \ref{fig:astroid}). Taking the Kerr axis along edge '1' of the triangle (Fig. \ref{fig:astroid}), we applied a $1~Hz$ sinusoidal magnetic field $H_{\Theta}$ of constant amplitude $H_{sat}$=410~Oe, at an angle $\Theta$ relative to edge '1'. Typical Kerr hysteresis loops are shown in Fig. \ref{fig:astroid}a, where the longitudinal magnetization is plotted versus the amplitude of the applied field.  They show a single transition for field angles $\left|\Theta\right|\leq75^{\circ}$ (e.g $\Theta = 0^{\circ}$, dashed line in Fig. \ref{fig:astroid}a), and two stepped transitions above this value (e.g $\Theta = 80^{\circ}$, solid line in Fig. \ref{fig:astroid}a). The lower transition tends to vanish altogether when the magnetic field is perpendicular to edge '1', for reasons made clear later in light of the simulations. The amplitude of the fields at transitions were recorded in $3^{\circ}$ steps between $+90^{\circ}$ and $-90^{\circ}$, and plotted against the field angle. The resulting polar plot (Fig. \ref{fig:astroid}b) shows a single point for angles $\left|\Theta\right|\leq75^{\circ}$ [modulus $180^{\circ}$], and two transitions for $\left|\Theta \right|>75^{\circ}$ [modulus $180^{\circ}$]. 

The loops taken with the Kerr axis along edge '1' reflect notable changes in longitudinal magnetization $M_{x}$ solely along edge '1'; equivalent measurements along the two other edges are therefore necessary to complete the astroid. The sample was therefore rotated by $120^{\circ}$ ($240^{\circ}$), and measured using edge '2' ('3') as the longitudinal Kerr axis. These measurements yielded identical first transitions, and second transitions for angles in the vicinity of $210^{\circ}$ [$180^{\circ}$] (opposite edge '2') and $-30^{\circ}$ [$180^{\circ}$] (opposite edge '3'). Overlaying the three sets of data gives the full astroid (Fig. \ref{fig:astroid}c). It is a twelve-pointed star presenting a six-fold symmetry. Maxima occur when the field direction is either along the edges of the triangle ($\Theta=0^{\circ}$ [$60^{\circ}$] at $H^{0}_{max}=153\pm5$~Oe), or along its axes of symmetry (second transitions at $\Theta=30^{\circ}$ [$60^{\circ}$] at $H^{30}_{max}=286\pm25$~Oe). Minima are present along the bisectors of the triangle at fields  $H^{30}_{min}=110\pm5$~Oe.

To complement these experiments, micromagnetic simulations were performed: $M_{s}=~800~kA/m$, $A_{ech}=13~pJ/m$ and $\alpha=0.5$, thickness $10$~nm, cell edge 2.5~nm for $e$ = 200, 300~nm, and 1~nm for $e$=100, 150~nm. We started by evaluating the influence of the physical rounding and size of the triangles on the respective stability of Y and buckle states.  Previous numerical work had established a stability phase diagram for the Y and buckle states as a function of triangle  size \cite{koltsov00}: for 10~nm thick non-rounded triangles, the Y state was found to be the most stable for $e\leq200$~nm. In our first set of simulations, the bitmap was not tilted as in Ref. \cite{koltsov00}, in order to allow full movement of spins along the base. The triangles  were first  initialized in a Y state perpendicular to edge '1', and then allowed to relax in zero field. In this way, stable Y  configurations were obtained in sharp triangles for  $e\leq200$~nm (Fig. \ref{fig:rounding}a), similar to Ref. \cite{koltsov00}. However, when including a finite rounding of the corners ($\oslash\geq10$~nm), no Y state could be stabilized at all for \textit{e}=100-300~nm. The most spectacular influence of the rounding is shown for \textit{e}=~100~nm in Fig. \ref{fig:rounding}. This is likely due to two concurrent effects: in sharp triangles, the exchange cost of spins in a buckle configuration is very high at the opposite corner, while at finite rounding the charge distribution in a buckle  is more favorable than in a Y state as the component of the magnetization normal to the three corners generates increasingly large stray fields with rounding. In the following simulations, a rounding of $70~nm$ was adopted, being closest to the experimental values.

\begin{figure}
	\centering
				\includegraphics[width=0.4\textwidth]{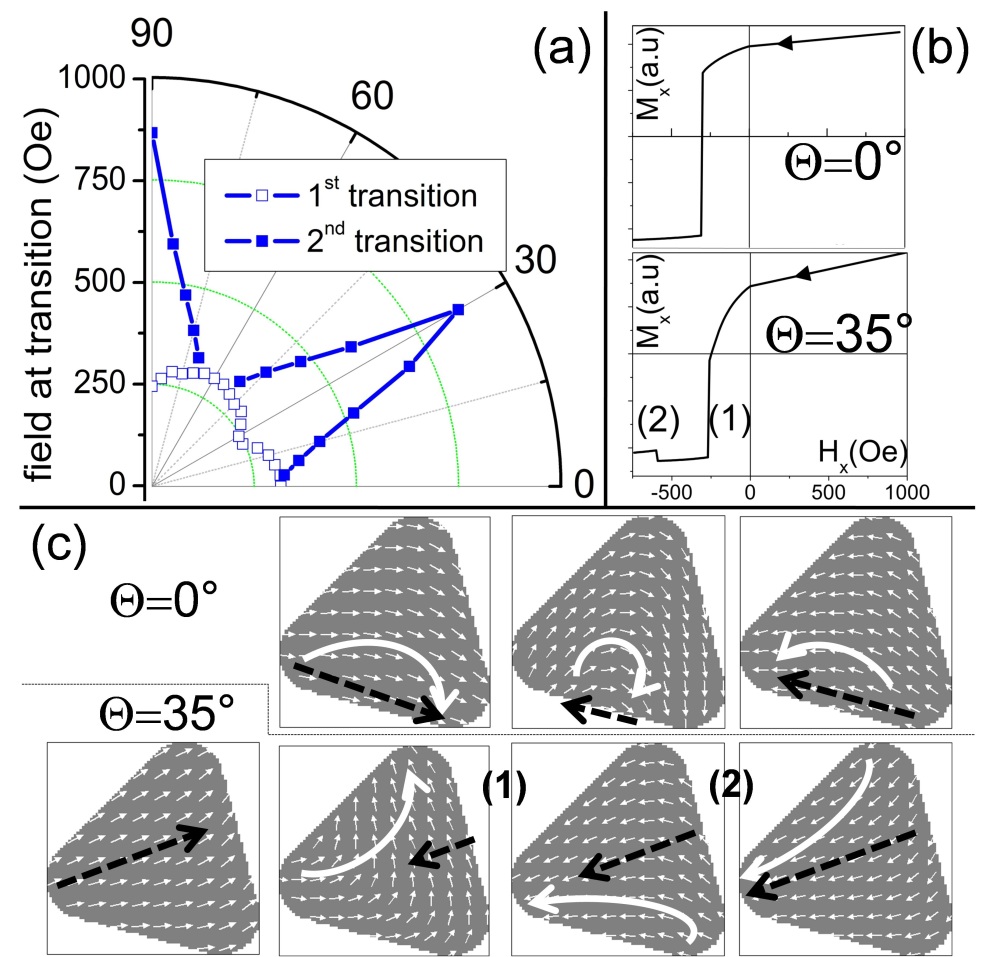}
	\caption{(color online) (a) Simulated switching astroid, open (closed) symbols correspond to first (second) transitions. (b) Magnetization $M_{x}$ plotted against $H_{x}$, projection along the triangle edge of the field applied along $\Theta = 0^{\circ}$ and $35^{\circ}$. (c) Simulated reversal mechanism for the field reversed along $\Theta$ (dashed line) in 300~nm triangles.}
	\label{fig:oommf}
\end{figure}

The switching astroid was then calculated by ramping the field from 0 to $-\vec{H}_{\Theta}$ along an angle $\Theta$ from the base of a $e$ = 300~nm triangle, after saturating along $\Theta$ at $H_{sat}$=1000~Oe. In this second set of simulations, the bitmaps were tilted by $15^{\circ}$ in order to spread pixellation effects along the three  edges during a complete magnetization reversal. Simulations were done between $0^{\circ}$ and $60^{\circ}$, in $5^{\circ}$ steps, and the fields inducing abrupt changes in the magnetization were plotted six times to give the full astroid, a quarter of which is shown in Fig. \ref{fig:oommf}a. Two types of magnetization reversal were found. For fields within $10^{\circ}$ of the $0^{\circ}$ [$60^{\circ}$] directions, the magnetization reversal occurred in a single transition, as shown in the $M(H)$ plot of Fig. \ref{fig:oommf}b (top). When the field is applied close to a triangle bisector on the other hand, the reversal takes place in two steps: a first transition at low fields, and a second transition occurring at increasingly high fields as $\Theta$ nears the   $30^{\circ}$ [$60^{\circ}$] directions (Fig. \ref{fig:oommf}b, bottom). With the magnetic field applied along $\Theta=0^{\circ}$ (Fig. \ref{fig:oommf}c, top), the magnetization naturally adopts a stable buckle state. As the field is reversed, a tight spin configuration curls along the edge, until the magnetization is abruptly reversed along the opposite $180^{\circ}$ buckle state.  When the field is applied along the $\Theta = 35^{\circ}$ axis on the contrary (Fig. \ref{fig:oommf}c, bottom), the magnetization does not adopt a Y configuration, but a strongly deformed $60^{\circ}$ buckle state. Upon reversal of the field, the magnetization first gradually flips to the $180^{\circ}$ buckle, reversing the sign of the longitudinal magnetization. Secondly, it abruptly reverses to the $-120^{\circ}$ buckle, seen as a slight decrease in  $M_{x}$, before eventually aligning itself along $\Theta=35+180^{\circ}$ at saturation. This second, higher field transition is therefore the one  observed by MOKE for fields within $15^{\circ}$ of the hard axes, and corresponds in effect to a 60$^{\circ}$ transition. Experimentally, it is most easily observed for fields in the vicinity of the perpendicular to the Kerr axis, $i.e$ for  $\left|\Theta\right|>75^{\circ}$. In the limiting case of a magnetic field absolutely perpendicular to the Kerr axis, the first transition is not observed by MOKE, because the initial and final states of transition (1) yield the same longitudinal magnetization, and only transition (2) is observed. The latter, being a minute rearrangement of spins, is however more difficult to observe, yielding larger error bars on the switching fields, and a slight asymmetry in the second transitions across the astroid.

The simulated switching astroid (Fig. \ref{fig:oommf}a) closely mimics the experimental one: maxima along  $0^{\circ}$ and $30^{\circ}$ ($H^{0}=305$~Oe and $H^{30}=865$~Oe), and minima along $30^{\circ}$  ($H_{min}=245$~Oe). The experimental switching fields were $0.5\pm0.1$ times lower than those obtained numerically, a regularly cited discrepancy \cite{himeno05,obrien09} due to the zero-temperature nature of the computation. The twelve peaks of the star can be intuited by a qualitative comparison to a Stoner Wohlfarth switching astroid \cite{stoner}. In the latter, each easy axis direction and each hard axis direction yields a local maximum in the astroid, leading to a 4 pointed star in the case of ellipses for instance. The Py triangles have 6 easy and 6 hard directions, thereby yielding a 12-pointed astroid. Note that the switching astroids complement the six-pointed anisotropy field polar plots of Ref. \cite{cowburn99}, obtained in Supermalloy triangles by modulated field magneto-optical anisometry (MFMA). These reflect the angular variation of the relative energies of buckle and Y states ($\Delta U=U_{buckle}-U_{Y})$, but do not highlight the different transitions occurring within the structure.

	Supported by the micromagnetic simulations, we can now affirm that the single step loops along the $0^{\circ}$ directions reflect the easy axis behavior of the buckle states, whereas no Y state could be stabilized  under linear fields. The easy axes lie along the edges of the triangle (6 buckle directions), and the hard axes along the perpendicular bisectors (6 Y directions).

We have studied in detail the six-fold anisotropy induced by shape symmetry in Py triangles, combining magneto-optical  experiments, and micromagnetic simulations. Very similar switching astroids were obtained by both methods, and explained by a detailed analysis of the different reversal mechanisms along hard  and easy axes of the triangle (resp. Y and buckle states). By including corner rounding in the simulations, we eliminated the stabilization conditions of the Y-state found in sharp triangles, and showed that only buckle states were stable down to 100~nm. Further work to be published on this system includes a systematic comparison of this behavior with a Stoner-Wohlfarth switching model.

We acknowledge E. R. Lewis for a critical reading of the manuscript.


\end{document}